\documentclass[twocolumn]{revtex4}

\usepackage{graphicx}
\usepackage{dcolumn}
\usepackage{amsmath}

\begin{document}

\title{Reply to "Comment on 'Isotope effect in multi-band and multi-channel attractive 
systems and inverse isotope effect in iron-based superconductors'"
}

\author{Takashi Yanagisawa$^{a,b}$, Kosuke Odagiri$^{a}$, Izumi Hase$^{a,b}$,
Kunihiko Yamaji$^{a,b}$,
Parasharam M. Shirage$^a$,
Yasumoto Tanaka$^{a}$, Akira Iyo$^{a,c}$, Hiroshi Eisaki$^{a,c}$}

\affiliation{$^a$Nanoelectronics Research
Institute, National Institute of Advanced Industrial Science and Technology (AIST),
Tsukuba Central 2, 1-1-1 Umezono, Tsukuba 305-8568, Japan\\
$^b$CREST, Japan Science and Technology Agency (JST), Kawaguchi-shi,
Saitama 332-0012, Japan\\
$^c$Transformative Research Project on Iron Pnictides (TRIP), JST, Sanbancho,
Chiyoda, Tokyo 102-0075 Japan
}


\maketitle

The Comment insists on the following: 
in our model it is assumed that the effective interactions
have specific energy ranges within the single band with a cutoff at $\omega_1$ for
the phononic part and a range from $\omega_1$ to $\omega_2$ in the AF channel.
Our reply is that 
we assume that $V_i({\bf k},{\bf k}')\neq 0$ if $|\xi_k|<\omega_i$ and 
$|\xi_{k'}|<\omega_i$, and otherwise $V_i({\bf k},{\bf k}')= 0$ ($i=1,2$), as stated
in our paper\cite{yan09}.
This is the model of BCS type with two attractive interactions, and this
assumption is the characteristic of the BCS approximation and is never unphysical.
The claim "the integration limits have been modified such that the AF channel
mediated pairing sets in where the ph-channel pairing terminates and is limited at 
an energy given by $\omega_j=\omega_{AF}$" in the Comment is completely wrong.
We also mention that
the eq.(1) in the Comment is incorrect since we cannot derive eq.(2) from eq.(1).
The eq.(2) is also wrong as shown below.
The eq.(5) in the Comment is also incorrect since this does not coincide with the
formula by Suhl et al. in the limit $\omega_{ph}=\omega_{AF}$\cite{suh59}.
We also show that the eq.(5) in the Comment is derived on the basis of an
unphysical model.

In the following, we describe the model and the method to solve the gap equation
in more detail.
Let us first consider the one-band and two-channel model with pairing interactions
$V_1$ and $V_2$.
We set $\omega_1<\omega_2$.
The interaction $V_2$ works in the range $0\leq |\xi_k|<\omega_2$.
In the one-band and two-channel model, the gap equation is
\begin{equation}
\Delta({\bf k})= \frac{1}{N}\sum_{{\bf k}'}\sum_{i=1}^2V_i({\bf k},{\bf k}')
\frac{\Delta({\bf k}')}{2E_{{\bf k}'}}\tanh\left(\frac{E_{{\bf k}'}}{2k_BT}\right),
\end{equation}
where $E_{\bf k}=\sqrt{\Delta({\bf k})^2+\xi_{{\bf k}}^2}$.
Within the BCS approximation, this is written as
\begin{eqnarray}
\Delta({\bf k})&=& N(0)\Big[ \int_{-\omega_1}^{\omega_1}d\xi_{k'}V_1({\bf k},{\bf k}')
+\int_{-\omega_2}^{\omega_2}d\xi_{k'}V_2({\bf k},{\bf k}')\Big]\nonumber\\
&\times&\frac{\Delta({\bf k}')}{2E_{{\bf k}'}}\tanh\left(\frac{E_{{\bf k}'}}{2k_BT}\right),
\end{eqnarray}
where $N(0)$ is the density of states at the Fermi level.
For $|\xi_{{\bf k}}|<\omega_1$, we obtain
\begin{eqnarray}
\Delta({\bf k})&=& (\lambda_1+\lambda_2)\int_{-\omega_1}^{\omega_1}d\xi_{k'}
\frac{\Delta({\bf k}')}{2E_{{\bf k}'}}\tanh\left(\frac{E_{{\bf k}'}}{2k_BT}\right)
\nonumber\\
&+& 2\lambda_2\int_{\omega_1}^{\omega_2}d\xi_{k'}
\frac{\Delta({\bf k}')}{2E_{{\bf k}'}}\tanh\left(\frac{E_{{\bf k}'}}{2k_BT}\right).
\label{gapeq1}
\end{eqnarray}
We defined $\lambda_i=N(0)\langle V_i\rangle_{FS}$ ($i=1,2$).
For $\xi_{{\bf k}}$ in the range of $\omega_1<|\xi_{{\bf k}}|<\omega_2$ where 
$V_1({\bf k},{\bf k}')=0$ and $V_2({\bf k},{\bf k}')\neq 0$, we have
\begin{equation}
\Delta({\bf k})= N(0)\int_{-\omega_2}^{\omega_2}d\xi_{k'}V_2({\bf k},{\bf k}')
\frac{\Delta({\bf k}')}{2E_{{\bf k}'}}\tanh\left(\frac{E_{{\bf k}'}}{2k_BT}\right).
\label{gapeq2}
\end{equation}
The assumption that $\Delta({\bf k})$ is constant being independent of ${\bf k}$ 
leads to a contradiction because
we obtain $k_BT_c=(2e^{\gamma}/\pi)\omega_2 e^{-1/\lambda_2}$ from eq.(\ref{gapeq2})
and
\begin{equation}
k_BT_c= \frac{2e^{\gamma}}{\pi}\omega_1^{\lambda_1/(\lambda_1+\lambda_2)}
\omega_2^{\lambda_2/(\lambda_1+\lambda_2)}\exp(-\frac{1}{\lambda_1+\lambda_2})
\end{equation}
from eq.(\ref{gapeq1}), where $\gamma$ is the Euler constant.
The latter coincides with $T_c$ of eq.(2) in the Comment.
These two $T_c$'s never coincide unless $\lambda_1=0$.
Hence, the $T_c$ of eq.(5) in the Comment is inconsistent and is wrong, 
and we must find a solution to 
gap equations that has the energy 
dependence\cite{bog59,mor62}.
Let us define $\Delta({\bf k})=\Delta_1$ for $|\xi_k|\leq\omega_1$ and
$\Delta({\bf k})=\Delta_2$ for $\omega_1<|\xi_k|\leq\omega_2$, then the critical
temperature $T_c$ is determined from
\begin{equation}
\Delta_1= (\lambda_1+\lambda_2)\Delta_1{\rm ln}\left(
\frac{2e^{\gamma}\omega_1}{\pi k_BT_c}\right)+\lambda_2\Delta_2{\rm ln}
\frac{\omega_2}{\omega_1},
\end{equation}
\begin{equation}
\Delta_2= \lambda_2\Delta_1\ln\left(\frac{2e^{\gamma}\omega_1}{\pi k_BT_c}\right)
+\lambda_2\Delta_2\ln\frac{\omega_2}{\omega_1}.
\end{equation}
The secular equation yields
\begin{equation}
k_BT_c= \frac{2e^{\gamma}\omega_1}{\pi}\exp\left(-\frac{1}{\lambda_1+\lambda_2^*}\right),
\end{equation}
where $\lambda_2^*=\lambda_2/(1-\lambda_2\ln(\omega_2/\omega_1))$.
This is the two-channel version of eq.(7) in Ref.\cite{yam87}.
This agrees with the formula by Morel and Anderson\cite{mor62} if $\lambda_2$ is the
Coulomb repulsive interaction with a negative sign.
The staircase gap function is a simplest one that gives a consistent solution to
gap equations with multi-cutoff energies.

The generalization to the two-band and two-channel model is quite straightforward.
We consider two bands denoted as $\alpha$ and $\beta$.
There are four interactions to be considered here: $V_{ph}^{\alpha\alpha}$, 
$V_{ph}^{\alpha\beta}$, $V_{AF}^{\alpha\alpha}$ and $V_{AF}^{\alpha\beta}$. 
They are intra- and inter-band
pairing interactions due to electron-phonon and antiferromagnetic interactions, respectively.
The coupled gap equations are
\begin{eqnarray}
\Delta^{\alpha}({\bf k})&=& -\frac{1}{N}\sum_{{\bf k}'}\sum_{i=ph,AF}\sum_{\mu=\alpha,\beta}
V_i^{\alpha\mu}({\bf k},{\bf k}')\frac{\Delta^{\mu}({\bf k}')}{2E^{\mu}_{{\bf k}'}}
\nonumber\\
&\times&\tanh\left(\frac{E^{\mu}_{{\bf k}'}}{2k_BT}\right),
\end{eqnarray}
and that for $\Delta^{\beta}$.
Here, $E_{{\bf k}}^{\mu}=\sqrt{\Delta^{\mu}({\bf k})^2+\xi_{{\bf k}}^2}$.
The energy range of the pairing interaction $V_{ph}^{\mu\nu}({\bf k},{\bf k}')$ is 
$0\leq |\xi_k|\leq\omega_{ph}$ and $0\leq |\xi_{k'}|\leq\omega_{ph}$,
and that of $V_{AF}^{\mu\nu}({\bf k},{\bf k}')$ is $0\leq |\xi_k|\leq\omega_{AF}$ and
$0\leq |\xi_{k'}|\leq\omega_{AF}$, for $\mu$, $\nu$= $\alpha$, $\beta$.  
We assume that $\omega_{ph}<\omega_{AF}$.
Outside of these ranges they vanish.
Then, the gap equations are written as
\begin{eqnarray}
\Delta^{\alpha}({\bf k})&=& -\sum_{\mu=\alpha,\beta}N^{\mu}(0)\Big[
\int_{-\omega_{ph}}^{\omega_{ph}}d\xi_{k'}V_{ph}^{\alpha\mu}({\bf k},{\bf k}')\nonumber\\
&+&\int_{-\omega_{AF}}^{\omega_{AF}}d\xi_{k'}V_{AF}^{\alpha\mu}({\bf k},{\bf k}')\Big]
\frac{\Delta^{\mu}({\bf k}')}{2E^{\mu}_{{\bf k}'}}
\tanh\left(\frac{E^{\mu}_{{\bf k}'}}{2k_BT}\right),\nonumber\\
\end{eqnarray}
and that for $\Delta^{\beta}$, where $N^{\mu}(0)$ is the density of states at the
Fermi level.

We define the coupling constants $\lambda_{i}^{\alpha\alpha}$
and $\lambda_i^{\alpha\beta}$ ($i$=ph, AF) similarly as in Ref.\cite{yan09}:
$\lambda_{AF}^{\mu\nu}=\langle N^{\nu}(0)V_{AF}^{\mu\nu}({\bf k},{\bf k}')\rangle_{FS}$ and
$\lambda_{ph}^{\mu\nu}=-\langle N^{\nu}(0)V_{ph}^{\mu\nu}({\bf k},{\bf k}')\rangle_{FS}$
for $\mu$, $\nu$= $\alpha$, $\beta$.
To obtain a self-consistent solution to gap equations, we set
$\Delta^{\mu}({\bf k})= \Delta^{\mu}_1$ for $0\leq |\xi_k|\leq\omega_{ph}$ and
$\Delta^{\mu}({\bf k})= \Delta^{\mu}_2$ for $\omega_{ph}< |\xi_k|\leq\omega_{AF}$,
for $\mu=\alpha$, $\beta$.
Then, the gap equations for $T_c$ are
\begin{eqnarray}
\Delta_1^{\alpha}&=&(\lambda_{ph}^{\alpha\alpha}-\lambda_{AF}^{\alpha\alpha})\Delta_1^{\alpha}
\ln\left(\frac{2e^{\gamma}\omega_{ph}}{\pi k_BT_c}\right)
-\lambda_{AF}^{\alpha\alpha}\Delta_2^{\alpha}\ln\frac{\omega_{AF}}{\omega_{ph}}\nonumber\\
&+& (\lambda_{ph}^{\alpha\beta}-\lambda_{AF}^{\alpha\beta})\Delta_1^{\beta}
\ln\left(\frac{2e^{\gamma}\omega_{ph}}{\pi k_BT_c}\right)
-\lambda_{AF}^{\alpha\beta}\Delta_2^{\beta}\ln\frac{\omega_{AF}}{\omega_{ph}},
\label{couple1}
\end{eqnarray}
\begin{eqnarray}
\Delta_2^{\alpha}&=& -\lambda_{AF}^{\alpha\alpha}\Delta_1^{\alpha}
\ln\left(\frac{2e^{\gamma}\omega_{ph}}{\pi k_BT_c}\right)
-\lambda_{AF}^{\alpha\alpha}\Delta_2^{\alpha}\ln\frac{\omega_{AF}}{\omega_{ph}}
\nonumber\\
&-&\lambda_{AF}^{\alpha\beta}\Delta_1^{\beta}
\ln\left(\frac{2e^{\gamma}\omega_{ph}}{\pi k_BT_c}\right)
- \lambda_{AF}^{\alpha\beta}\Delta_2^{\beta}\ln\frac{\omega_{AF}}{\omega_{ph}},
\label{couple2}
\end{eqnarray}
and those for $\Delta_1^{\beta}$ and $\Delta_2^{\beta}$.
We set $\lambda_{ph}^{\alpha\beta}=\lambda_{ph}^{\beta\alpha}$ and
$\lambda_{AF}^{\alpha\beta}=\lambda_{AF}^{\beta\alpha}$ since the mutual pair
transfer interactions are the same between bands $\alpha$ and $\beta$.
For simplicity, we assume that 
$\lambda_{ph}^{\alpha\alpha}=\lambda_{ph}^{\beta\beta}$,
$\lambda_{AF}^{\alpha\alpha}=\lambda_{AF}^{\beta\beta}$ and $N^{\alpha}=N^{\beta}$.

Let us first consider the solution to this coupled equation, with the $s_{\pm}$ symmetry 
satisfying
$\Delta_i^{\alpha}=-\Delta_i^{\beta}$ ($i=$1, 2).
From eq.(\ref{couple2}), the ratio 
$y\equiv \Delta_2^{\alpha}/\Delta_1^{\alpha}=\Delta_2^{\beta}/\Delta_1^{\beta}$ is
written as
\begin{equation}
y= \frac{\lambda_{AF}^{\alpha\beta}-\lambda_{AF}^{\alpha\alpha}}
{1+(\lambda_{AF}^{\alpha\alpha}-\lambda_{AF}^{\alpha\beta})\ln(\omega_{AF}/\omega_{ph})}
\ln\left(\frac{2e^{\gamma}\omega_{ph}}{\pi k_BT_c}\right).
\end{equation}
We obtain the critical temperature, by substituting $y$ into eq.(\ref{couple1}),
\begin{equation}
k_BT_c= \frac{2e^{\gamma}}{\pi}\omega_{ph}\exp
\left(-\frac{1}{\lambda_{ph}+\lambda_{AF}^*}\right),
\end{equation}
where $\lambda_{ph}=\lambda_{ph}^{\alpha\alpha}-\lambda_{ph}^{\alpha\beta}$,
$\lambda_{AF}=\lambda_{AF}^{\alpha\beta}-\lambda_{AF}^{\alpha\alpha}$, and
$\lambda_{AF}^*=\lambda_{AF}/(1-\lambda_{AF}\ln(\omega_{AF}/\omega_{ph}))$.
It is obvious that we cannot obtain a consistent solution if we assume that $y=1$, that is,
$\Delta^{\mu}$ are constant.
The above derivation of $T_c$ is very simple and natural in the BCS approximation,
 and thus we can discuss the isotope effect on the basis of this formula\cite{shi09}.
The isotope coefficient $\alpha$ is derived as
\begin{equation}
\alpha= \frac{1}{2}\Big[ 1-\left(\frac{\lambda_{AF}^*}{\lambda_{ph}^{\alpha\alpha}
-\lambda_{ph}^{\alpha\beta}+\lambda_{AF}^*}
\right)^2\Big].
\end{equation}
The physics that leads to negative $\alpha$ is very clear.
In the pairing state with $s_{\pm}$ symmetry, the negative $\alpha<0$ occurs if
the inter-band electron-phonon coupling $\lambda_{ph}^{\alpha\beta}$ is larger than
the intra-band one $\lambda_{ph}^{\alpha\alpha}$.
Thus, the inverse isotope effect stems from the inter-band electron-phonon interaction.

Second, let us investigate the $s_{++}$ state.
In this case, we adopt $\Delta_i^{\alpha}=\Delta_i^{\beta}$ ($i$=1,2).
We obtain, from eq.(\ref{couple2}),
\begin{equation}
y= -\frac{\lambda_{AF}^{\alpha\alpha}+\lambda_{AF}^{\alpha\beta}}
{1+(\lambda_{AF}^{\alpha\alpha}+\lambda_{AF}^{\alpha\beta})\ln(\omega_{AF}/\omega_{ph})}
\ln\left(\frac{2e^{\gamma}\omega_{ph}}{\pi k_BT_c}\right).
\end{equation}
The substitution of $y$ to eq.(\ref{couple1}) yields
\begin{equation}
k_BT_c= \frac{2e^{\gamma}\omega_{ph}}{\pi}\exp\left(-\frac{1}{\lambda_{ph}^+-
(\lambda_{AF}^+)^*}\right),
\end{equation}
where $\lambda_{ph}^+=\lambda_{ph}^{\alpha\alpha}+\lambda_{ph}^{\alpha\beta}$,
$\lambda_{AF}^+=\lambda_{AF}^{\alpha\alpha}+\lambda_{AF}^{\alpha\beta}$ and
$(\lambda_{AF}^+)^*=\lambda_{AF}^+/(1+\lambda_{AF}^+\ln(\omega_{AF}/\omega_{ph}))$.
Since 
$d\ln(k_BT_c)/d\ln\omega_{ph}=1-[(\lambda_{AF}^+)^*/(\lambda_{ph}^+-(\lambda_{AF}^+)^*)]^2$,
the isotope coefficient is
\begin{equation}
\alpha= \frac{1}{2}\Big[ 1-\left(\frac{(\lambda_{AF}^+)^*}
{\lambda_{ph}^+-(\lambda_{AF}^+)^*}\right)^2\Big].
\end{equation}
This gives the positive isotope effect $\alpha>0$, except the case where
$(\lambda_{AF}^+)^*<\lambda_{ph}^+<2(\lambda_{AF}^+)^*$.
Hence, the isotope effect is probably normal in the $s_{++}$-pairing state.

The critical temperature $T_c$ in eq.(5) of the Comment may be derived from
the following coupled equation,
\begin{equation}
\Delta^{\alpha}=\lambda_{AF}^{\alpha\alpha}\Delta^{\alpha}\ln\frac{C\omega_{AF}}{k_BT_c}
+\lambda_{AF}^{\alpha\beta}\Delta^{\beta}\ln\frac{C\omega_{AF}}{k_BT_c},
\end{equation}
\begin{equation}
\Delta^{\beta}=\lambda_{ph}^{\beta\beta}\Delta^{\beta}\ln\frac{C\omega_{ph}}{k_BT_c}
+\lambda_{ph}^{\beta\alpha}\Delta^{\alpha}\ln\frac{C\omega_{ph}}{k_BT_c}.
\end{equation}
In fact, if we assume that the gap functions $\Delta^{\mu}$ are constant, we obtain
$T_c$ in the Comment (with some corrections).
The model that gives this coupled equation is, however, unphysical because the
Hamiltonian is inevitably not hermitian.  Thus, it is not proper to apply this model to 
real Fe pnictides.


\begin{thebibliography}{}
\bibitem{yan09}T. Yanagisawa, K. Odagiri, I. Hase, K. Yamaji, P. M. Shirage,
Y. Tanaka, A. Iyo, H. Eisaki: J. Phys. Soc. Jpn. 78 (2009) 094718.
\bibitem{suh59}H. Suhl, B. T. Matthias, and L. R. Walker: Phys. Rev. Lett. 3 (1959)
552.
\bibitem{bog59}N. N. Bogoliubov, V. V. Tolmachev and D. V. Shirkov:
Fort. der Phys. 6 (1958) 605. 
English translation in {\em  Theory of
Superconductivity} edited by N. N. Bogoliubov (Gordon and Breach
Science Publishers, 1968).
\bibitem{mor62}P. Morel and P. W. Anderson: Phys. Rev. 125 (1962) 1263.
\bibitem{yam87}K. Yamaji: Solid State Commun. 61 (1987) 413.
\bibitem{shi09}P.M. Shirage, K. Kihou, K. Miyazawa, C.-H. Lee, H. Kito,
H. Eisaki, T. Yanagisawa, Y. Tanaka, A. Iyo: Phys. Rev. Lett. 103 (2009) 257003.
 

\end{thebibliography}
\end{document}